# Data warehouse on Manpower Employment for Decision Support System

Amro F. ALASTA, and Muftah A. Enaba

***Abstract*—**Since the use of computers in business world, data collection has become one of the most important issues due to the available knowledge in the data; such data has been stored in database. Database system was developed which led to the evolvement of hierarchical and relational database followed by Standard Query Language (SQL). As data size increases, the need for more control and information retrieval increase. These increases lead to the development of data mining systems and data warehouses.

This paper focuses on the use of data warehouse as a supporting tool in decision making. We to study the effectiveness of data warehouse techniques in the sense of time and flexibility in our case study (Manpower Employment). The study will conclude with a comparison of traditional relational database and the use of data warehouse.

The fundamental role of data warehouse is to provide data for supporting decision-making process. Data in data warehouse environment is multidimensional data store. We can simply say that data warehouse is a process not a product, for assembling and managing data from various sources for the purpose of gaining a single detailed view of part or all an establishment. The data warehouse concept has changed the nature of decision support system, by adding new benefits for improving and expanding the scope, accuracy, and accessibility of data.

The warehouse is the link between the application and raw data, which is scattered in separate database but now is unified.

The objectives of this work are to study the impact of using data warehouse on Manpower Employment Decision Support System, in the sense as far as the data quality concern. We will focus on the benefits gained from using data warehouse, and why it is more powerful than the use of traditional databases in decision making. The case study will be the Libyan national manpower employment agency. The data warehouse will collect database scattered from different sources in Libya in order to compare the performance and time.

## I. Introduction

SINCE the late 1960s, the area of Database Management System (DBMS) has emerged in response to the need of many organizations to manage and benefit from these huge amounts of data, which have been collected and generated by these organizations.

The concept of tabular oriented relational database was introduced in the early 70s by Dr. Ted Cod. The relational database model has received much attention and developments due to its simple mathematical basis (the set theory). Commercially viable, relational database management systems were available in the market by early 80s. Although, in the early 1980s, most of the commercial database systems were based on relational models, several alternatives in database models were also proposed. One of those alternatives for relational database was the semantic data model. Another alternative is the Object Orientation (OO) model, the purpose behind both the development of semantic data models and the development OO models is to model the real world as closely as possible. In OO data modeling, each real world entity of problem domain is represented by a set of objects with relations and operations. Each object consists of part of objects or sub-objects that relates objects to each other (relation representation).

In the early 1990, relational database management systems were more popular than hierarchal and network database management systems. Some of the increased advantages of the relational database management systems were its functionality and flexibility and the use of cache up in performance. In current database management systems, object oriented techniques become more popular because of its encapsulation of the data and the functions being performed on these data.

Recently, advances in technology have been revealing new applications of database systems, such as pictures, video clip, and sound message; can now be stored by multimedia database. In addition, maps, weather data, and satellite images can be stored and analyzed by Geographic Information System (GIS).

As data size increases, the needs for more control and information retrieval also have increased. These increases have led to the development of Data Warehouses (DW), Data Mining (DM) systems and Knowledge Discovery in Database (KDD) systems. "Data warehouse (DW) and On Line Analytical Processing (OLAP) system are used in many companies to extract and analyze useful information from very large database for decision making. Real-time and active database technologies are used in controlling industrial and manufacturing processes. Furthermore, database search techniques are being applied to the World Wide Web (WWW) to improve the search for information that is needed by users browsing through the internet." [1]. New generation of integrated information systems have been appealed to

Amro F. ALASTA, Faculty of Science - Misurata University, Zliten, LIBYA. Email: Amr_hard@yahoo.com

Muftah A. Enaba, Faculty of Education -University, Misurata, LIBYA. Email: eniba_muf@yahoo.com.

computer users since the year of 2000.

There are many definitions of data mining, one of the most common definitions describe data mining as, "data mining is the process of discovering interesting knowledge from large amounts of data stored either in databases, data warehouses, or other information repositories." [5]. Another definition of data mining is given in [3], states it as; the search for relationships and global patterns that exist in large databases that are hidden among the vast amount of data, such as the relationship between patient data and their medical diagnosis.

All above definitions and other ones put stress on the discovery of relationships, patterns and trends in vast amount of data in figure.

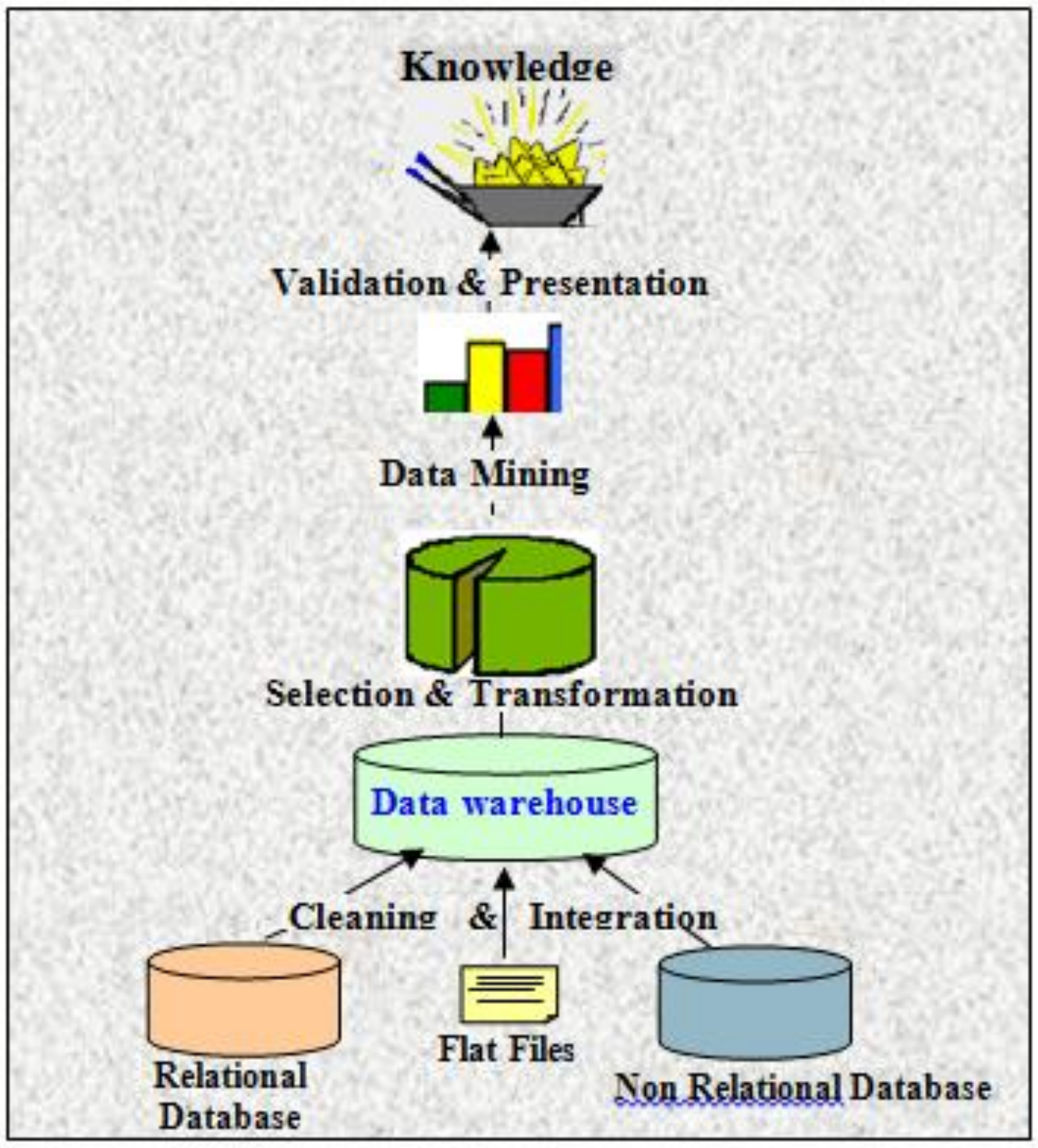

Fig.1 Data mining as a process of knowledge discovery

## II. DATA WAREHOUSE ARCHITECTURE

"Many researchers and practitioners share the understanding that a data warehouse (DW) architecture can be formally understood as layers of materialized views on top of each other. DW architecture exhibits various layers of data in which data from one layer are derived from data of the lower layer." [17]. Here, we will illustrate the layers that constitute a data warehouse and as depicted in figure-2.. The lowest layer of the data warehouse architecture is called the data sources layer, which usually consists of the operational databases. This layer may consist of structured, unstructured or semi-structured data stored in files or other storage system. The data in this layer is extracted to create the data warehouse.

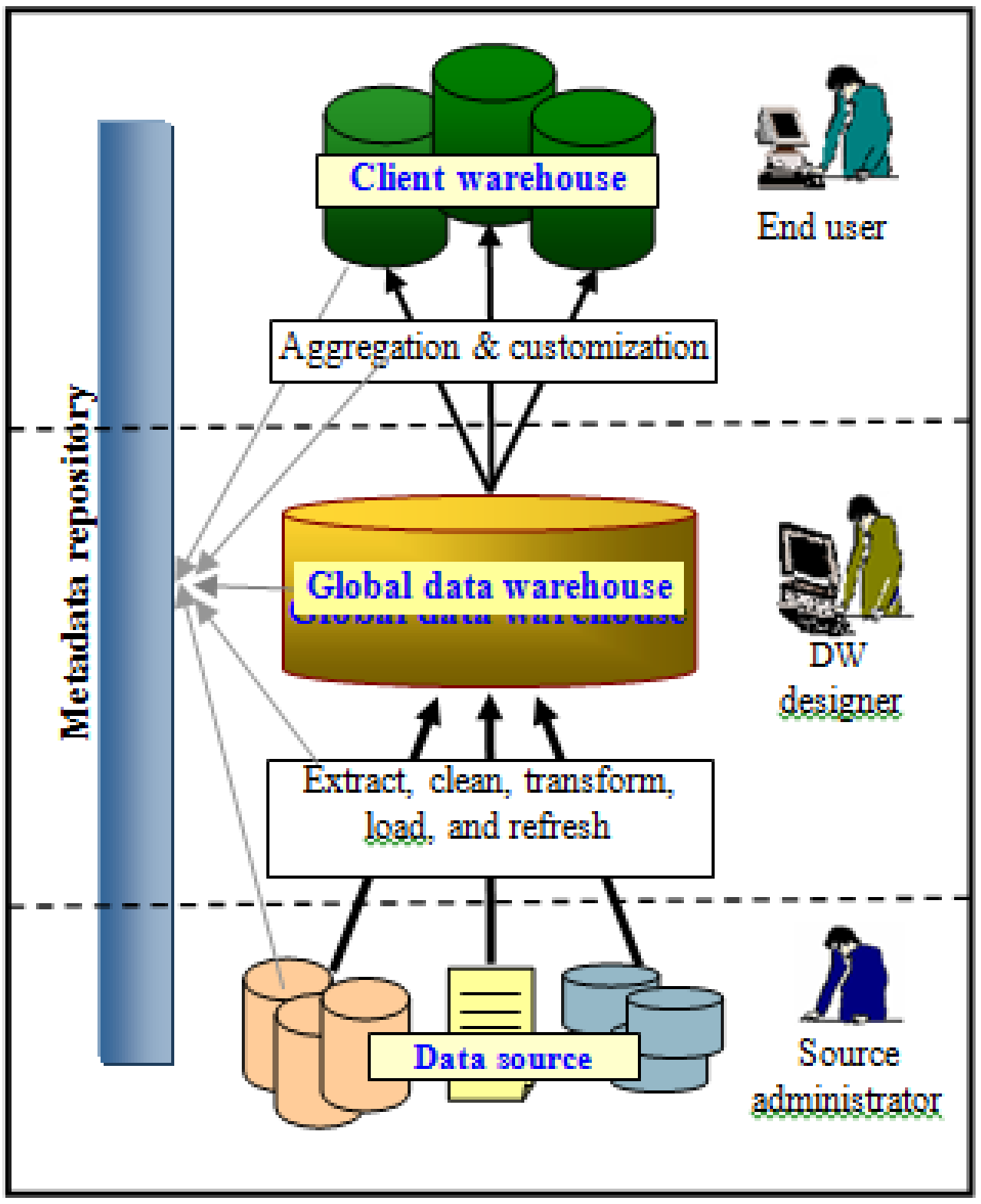

Fig. 2 Illustrate the Data Warehouse Architecture

The middle layer of the architecture is the global data warehouse. In this layer a historical record of data is stored after being resulted from some operations such as: transformation, integration, and aggregation of detailed data found in the data sources. The data warehouse is populated with clean and homogeneous data.

## III. ON LINE TRANSACTION PROCESSING (OLTP) VS. ON LINE ANALYTICAL PROCESSING (OLAP).

The purpose of On Line Transaction Processing (OLTP) systems is to allow high concurrency between uses which make it possible for many users to access the same data at the same time. As the name implies, these systems allow transactions to be processed against the data. In other words, these systems control the changes of the data due to some operations such as: insertion, update and deletion during business processes. Figure-3. Depicts a basic OLTP system:

The figure shows that numerous client applications can access the database to get the needed pieces of information. The broken lines between the client applications and the DBMS symbolized that these connections can physically be implemented in many different ways.

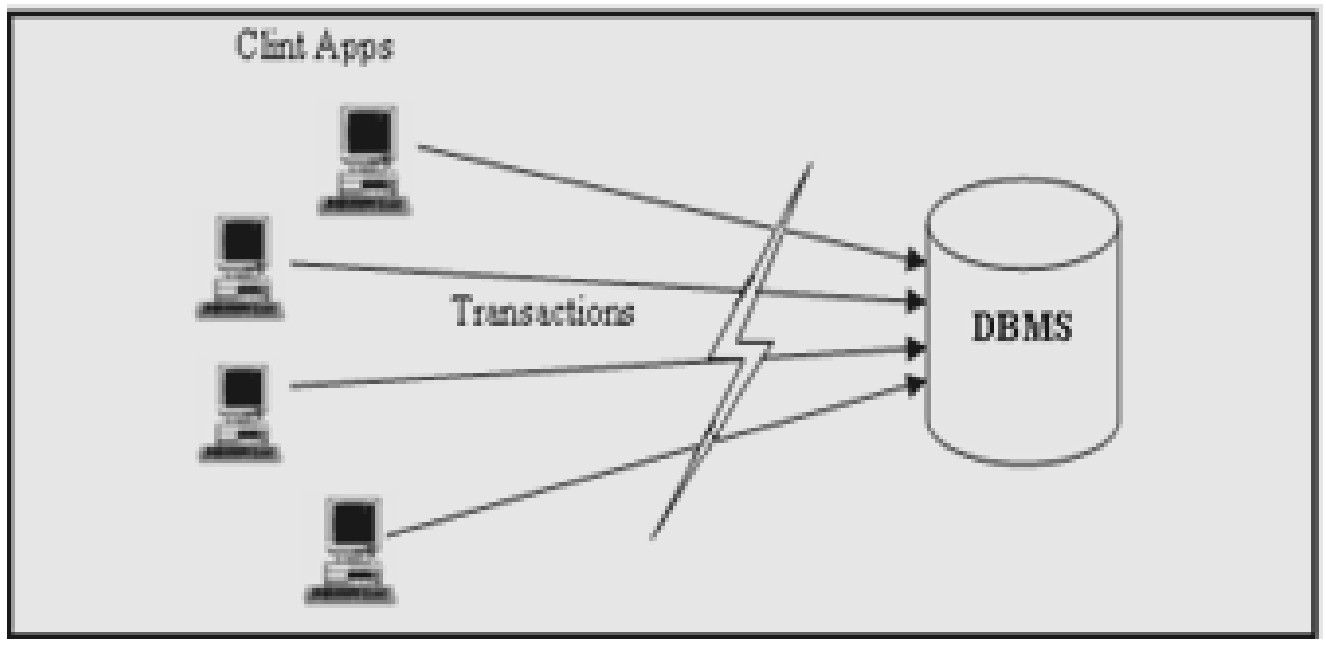

Fig. 3 Depicts a basic OLTP system.

## IV. OLD SYSTEM PROBLEMS

The job seekers apply for the suitable job according to their specialization (major), education level and the jobs offered in their city of residence. The applicants have to apply for job in any of the (GPCM) centers in each city. The General People's Committee of Manpower is doing its best in manually form to collect the reports produced by the computerized systems in each city. However each city in Libya has its own computer system to help in un employing people, but there are two main problems with these systems which are:

1. The reporting module takes a lot of time to extract some statistical reports about the applicants.
2. The system in each city works in an individual manner with its own database that differs from the other databases in other cities, which makes it very difficult to extract, and collect all the statistical reports needed for all cities together.

According to the currently used information system in the General People's Committee of Manpower (GPCM) and based on our study of a sample of three centers in Tripoli, Misurata and sirite popularities, we realized that each center has its own separated database that differs in structure, coding, data types and field lengths, from the ones used in other centers. For example in Tripoli the files used are flat files data format, in Misurata a relational database is used, while in sirite non-relational database is used. Therefore we have realized that the current system of the General People's Committee of Manpower has the following drawbacks:

1. The administration of data is complex.
2. There is data inconsistency.
3. Solving inconsistencies is expensive and too slow.
4. The needed informational data for the decision support takes long time to acquire.
5. The decision support needs an informational data not a raw data.
6. The traditional decision support environment has failed to provide complete, accurate, integrated, and timely information to the secretary of manpower

These deficiencies motivated us to investigate the idea of using data warehouse in the decision support sector of the General People's Committee of Manpower.

## V. OLTP SYSTEM STRUCTURE

Each local Secretary of Manpower in each city in Libya has its own database, each of which contains various database tables. As an example, table-1 depicts Misurata Citydatabase tables.

TABLE I
MISURATA CITY DATABASE

| Table Name | Table Description |
|---|---|
| Applicant | This table contains personal data of the applicant such as; ID-number, name address, sex. |
| Specialty | This table contains the information related to the specialty (major) of the applicant. |
| Job-Group | Contains the attributes that describe the job-group specified in each city |
| Sector | Contains the attributes that describe the sector to which, applicants were directed to. |
| Moahel | Contains the attributes that describe the education qualification of the applicant. |
| Education-level | Contains the attributes that describe the education level of applicant, like primary preparatory, secondary, … |
| Mothamer | Contains the attributes that describe the congresses in each city. |
| Service | Contains the attributes that are related to the military service record for males or national service for females. |

## VI. METHODOLOGY

In order to build the required data warehouse, we started by collecting information from scattered databases in the above mentioned three cites targeted in our study and store the data under a unified schema. As we will illustrate bellow, we constructed data warehouse via a process of data cleaning, transformation, integration, reduction, loading, and periodic data refreshing.

The design of the relation database requires the use of Entity Relationship (ER) model, which is appropriate for On-Line Transactional Processing (OLTP). Data warehouse design requires subject schema that is appropriate for On Line Analytical Processing (OLAP). "The most popular data model for data warehouse is a multidimensional model. Such model can exist in the form of star schema, a snowflake schema, or a fact constellation schema." [5]. According to [11], most data warehouses use a star schema to represent the multidimensional data model, so we adopt this type of schema in the design of the General People's Committee of Manpower data warehouse.

## VII. DATA PREPROCESSING

Typically, the Manpower databases, as an example of real-world databases that are highly susceptible to noise, missing, and inconsistent data due to their normally huge size. So it is better to improve the quality of the data by preprocessing and in turn this will improve the mining task.

During the implementation of the GPCM data warehouse, we were faced with two main problems. The first problem is the redundancy of the applicant's records, where applicants have applied in more than one city. We have to overcome this problem by the use of cleansing procedure in our data warehouse tools that deletes any redundant records. The second problem is that of entry data discrepancy. Because there are about 25 centers for data entry and each operator has its own style for data entry.

We have solved the above mentioned problems by some preprocessing procedures in our data warehouse tools to perform cleansing, integration, transformation and data reduction.

## VIII. DATA CLEANING

There are many possible reasons for noisy data, at most human error occurring at data entry, the data cleaning routines work to clean the data by filling in missing values using a global constant, a most probable value, or using an attribute mean for numeric values. We used a global constant approach, which is replacing all the missing value by the same constant. Even though this approach is simple and easy to implement, it is not recommended because it might affect the mining process because it could be mistakenly taken to be as an interesting concept.

## IX. DATA TRANSFORMATION

Data transformations involve multiple techniques like, smoothing, aggregation, generalization, normalization and attribute construction. In our implementation, generalization is used to replace low-level data by higher-level concepts by the use of concept hierarchies. For example categorical address in which districts are generalized to high-level concept “congress”, as depicted in figure-4.

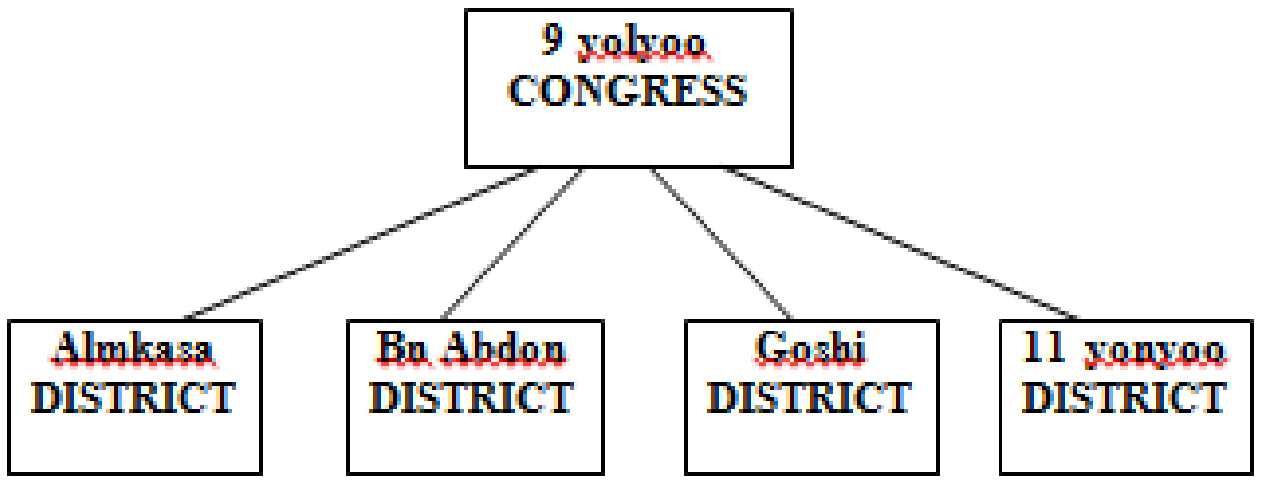

Fig. 4 A concept hierarchy for attribute address in Misurata city

## X. DATA REDUCTION

The manpower's data is huge and complex, so it takes a long time to be processed, that makes it impractical for data mining. So, it is important to reduce the data size and remove the irrelevant attributes for the mining process, in a way that preserves the integrity of the original data. To reduce the data, number of data reduction techniques could be applied such as; dimension reduction, where the irrelevant or redundant attributes are removed. Another data reduction technique which is known as data cube aggregation might be also used where aggregation operations are applied to the data and the result is stored in a multidimensional data cube. For example, the manpower's database contains data for directed applicants for different jobs per quarter for the years from 2000 to 2006. However, sometimes it’s of more interest if the total directed applicant by year rather than by quarter, so the aggregated totals are resulted in a smaller volume without losing any information. This type aggregation is illustrated in figure-5.

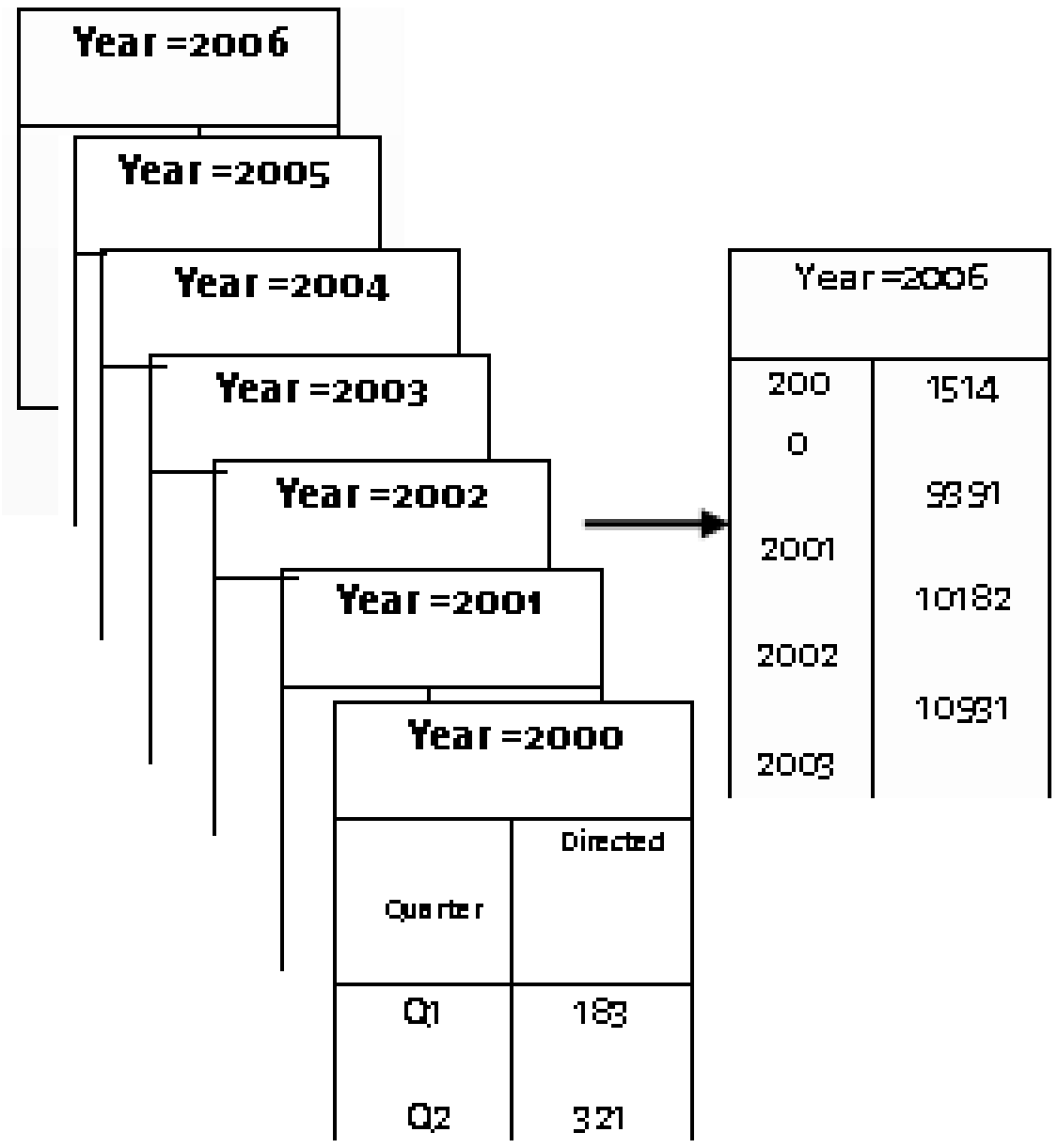

Fig. 5 Aggregation is illustrated

Concept hierarchy technique is also used in data reduction. For example in our case study the values of the attributes city, congress, district form a concept hierarchy with multiple levels as depicted in figure-6.

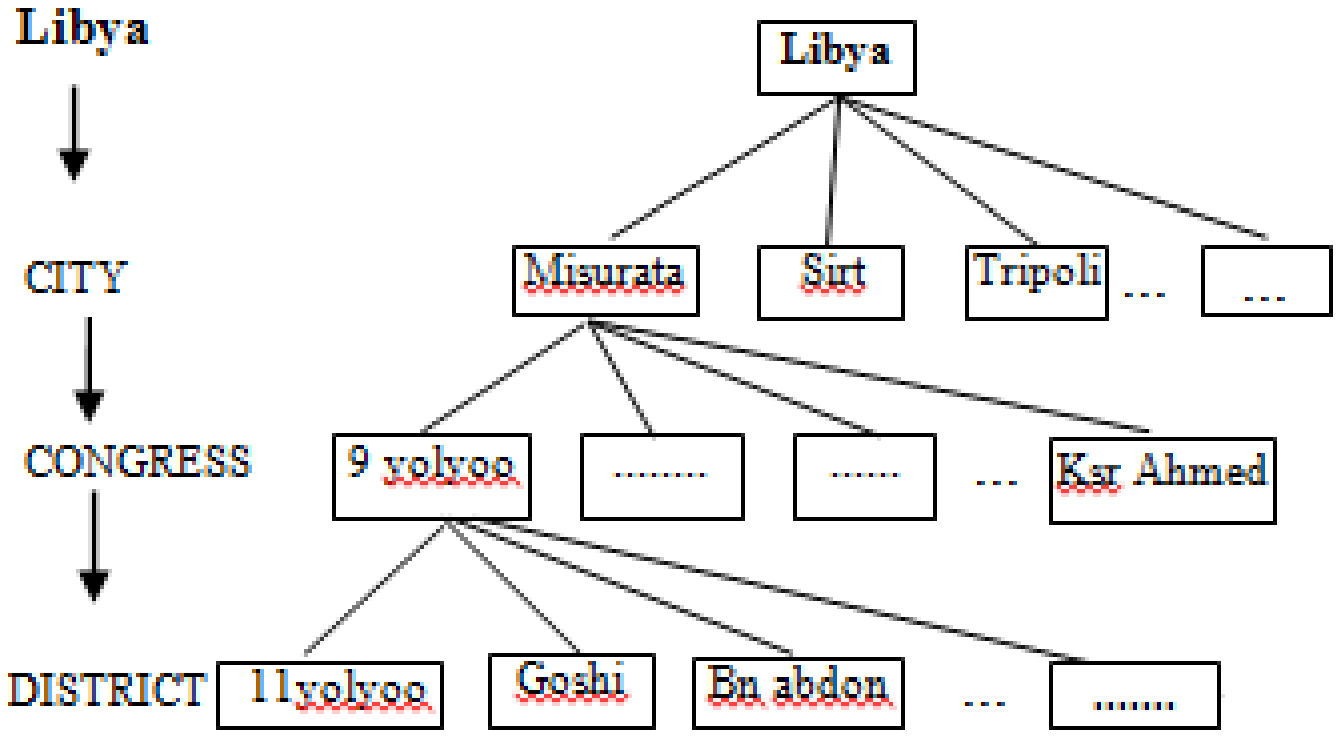

Fig. 6: A concept hierarchy for attribute address in

## XI. BUILDING GPCM DATA WAREHOUSE

In the process of building the General People's Committee of Manpower data warehouse, it is sufficient to use star schema model, which containts a single fact table and a number of dimensional tables. Table II and table III, depicts the structure of the fact table and the dimension tables respectively.

TABLE II STRUCTURE OF THE FACT TABLE

| **Attribute name** | **Description** | **Remark** |
|---|---|---|
| City_id | Each city has an unique number | Key |
| Sector_id | Each sector has an unique number | Key |
| EduLevel_id | Each education level has an unique number | Key |
| Cong_id | Each congress has an unique number | Key |
| Service_id | Each military service has an unique number | Key |
| Time_id | Each time unit has an unique number | Key |
| Total number of applicant | Number of all applicants | Measure |
| Number of seekers | Number of all seekers | Measure |
| Number of directed | Number of all directed | Measure |
| Education level | Number of applicants for specific education level | Measure |
| Military service | Number of applicants for a specific military service | Measure |

The fact table contains a number of measures that constitute the output. These measures are:

- Total number of applicants represents the number of applicants that can be reported accurately in a very short time.
- Number of seekers represents the number of applicants that seek to be employed for each city and/or for a specific period of time.
- Number of directed applicants represents the number of applicants that have been directed to a specific job. The number of directed applicants can be tabled in many ways such as; the number of directed applicants to a specific sector or in a given city or in a given period of time.
- Education level count represents the number of applicants in each education level.
- Military service count represents the number of applicants in each military service category in a given city or for all cities in a given period of time.

TABLE III STRUCTURE OF THE DIMENSION TABLE

| Dimension Name | Dimension Description |
|---|---|
| City | Contains the attributes that describe the city |
| Descen_Sector | Contains the attributes that describe the sector to which the applicant is directed. |
| Education_level | Contains the attributes that describe the education level |
| Congress | Contains the attributes that describe the congress in each city |
| Service | Contains the attributes that describe the military service for male or social service for female |
| Time | Contains the attributes that describe the time dimension, each year consists of 4 quarters. |

## XII. GPCM MULTIDIMENSIONAL DATA MODEL

The GPCM multidimensional data model can be visualized as data cube with several dimensions, as depicted in figure- 7. The city dimension consists of the cities in Libya in our case study. The time dimension divides the year into four quarters (Q1, Q2, Q3 and Q4). The education level dimension is divided into six educational levels.

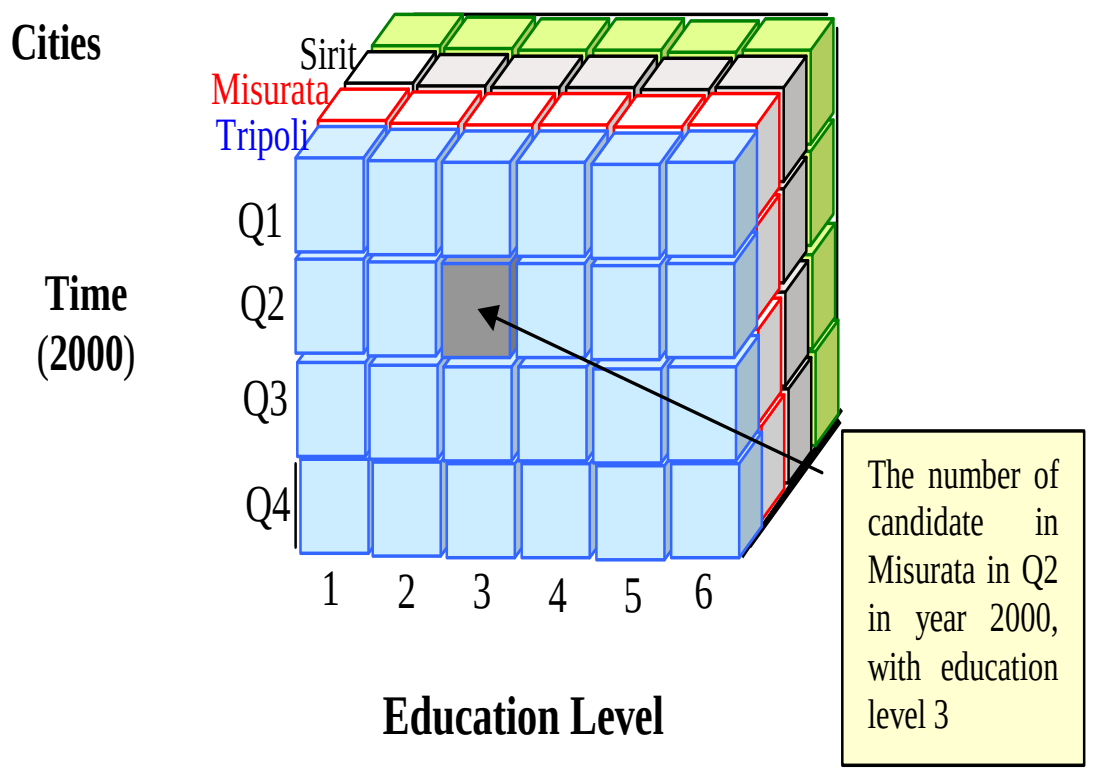

Fig. 7 GPCM Multidimensional data model

Next figure-8 depicts the total number of job seekers, according to their preferred sectors in all cities from the year of 2000 to the year of 2006.

Figure-9, represents a comparison between seekers for jobs and the directed applicants to different sectors from the year 2000 to 2006 in all cities.

The previous graphical representations and reports are the results of our data warehouse system which will help the decision makers in taking the right decisions in the right time with no delay according to the new information available.

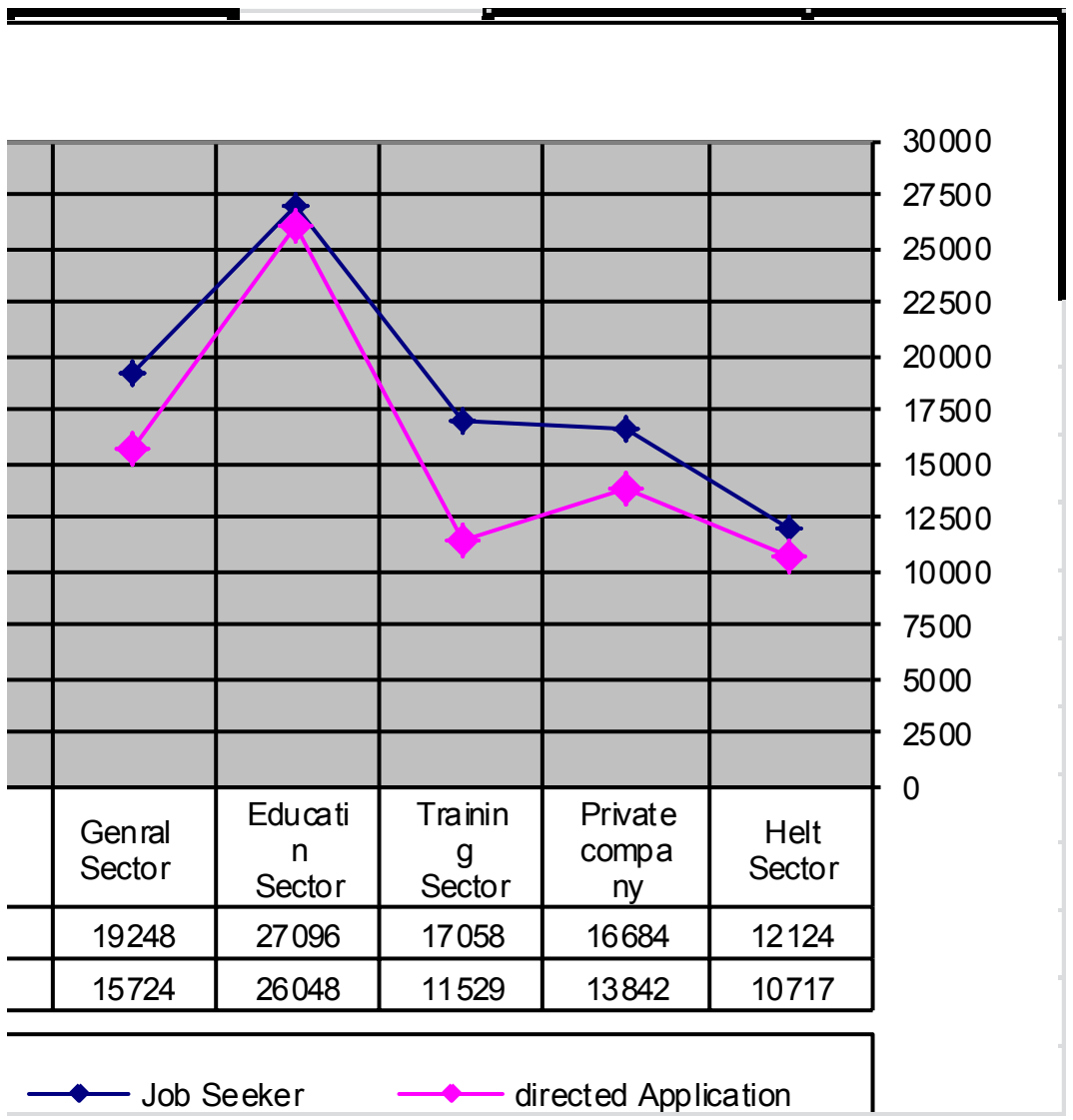

Fig. 8 Graphical representation for the number of applicants and directed applicants in all cities from the year 2000 to 2006

## XIII. Conclusion

"As our world is now in its information era, a huge amount of data is accumulated everyday. A real universal challenge is to find actionable knowledge from a large amount of data. Data mining is an emerging research direction to meet this challenge. Many kinds of knowledge (patterns) can be mined from various data", [15]. The objectives of this work are to study the impact of using data warehouse on a huge amount of data in Manpower Employment Decision Support System in which the data quality are concerned. In our research, we compared the data warehouse with the OLTP system to know the issues and impacts of using the data warehouse system on the information system. In fact, we found that we cannot completely replace the information system with the data warehouse but the data warehouse is just for helping the information system in the process of decision-making. We measured the reporting performance between the two systems and found that the data warehouse is more powerful than the ordinary old system. The old system has its features, which the data warehouse cannot do it (as an example add a record, update a record or delete a record), this is because it is not its functions. We cannot use the old system instead of the data warehouse especially in decision support reports or multidimensional reports, which need a lot of time from the Information System staff, and at the end, these reports are static reports not dynamic ones.

We have accomplished the objectives of the study by focusing on the benefits gained from using data warehouse, and why it is more powerful than the use of traditional databases in decision-making. The case study tackled the impact of using the data warehouse on employing people in People's Committee of Manpower .We have analyzed and designed it by V.B studio 6.0, and SQL services programming languages, in order to compare the performance and time. We tested our system with real, synthetic and scattered data that was collected from different sources in Libya with different types (access database, DBF files and flat file) and sizes (15.56 MB, 6.85 MB and 28.1 MB).

The problems that face the old system users, which we mentioned in section 4.1 motivated us to use data warehouse system in the decision support organization. The advantages of this system were shown in section 4.4, where the resulted reports and figures emphasize those advantages and the ability of such system to help decision making which is not easy to achieve using the old system. The reports and figures produced by the presented system show how clear the overall situation of the employment process. So decision makers can take the right decisions in the right time with no delay according to the new information available.